\documentclass[useAMS,a4paper]{mn2e}
\usepackage{graphicx}
\usepackage{epstopdf}
\usepackage{color}

\title[sSFR: spheroids and discs]{The two regimes of the cosmic sSFR evolution
are due to spheroids and discs }

\author[A. Pipino, F. Calura, \& F. Matteucci]{A. Pipino\thanks{pipinoa@phys.ethz.ch}$^1$, F. Calura$^2$
\& F. Matteucci$^{3,4}$ \\
$^1$ Institut f\"ur Astronomie, ETH Zurich, Wolfgang-Pauli-Str. 27, 8093, Zurich, Switzerland \\
$^2$ INAF Osservatorio Astronomico di Bologna, \\
$^3$ Dipartimento di Fisica, Sez. Astronomia, Universit\'a di Trieste, via G.B.Tiepolo 11, 34100, Trieste, Italy \\
$^4$ INAF Osservatorio Astronomico di Trieste, via G.B.Tiepolo 11, 34100, Trieste, Italy}

\long\def\symbolfootnote[#1]#2{\begingroup%
\def\thefootnote{\fnsymbol{footnote}}\footnote[#1]{#2}\endgroup}

\begin{document}

\date{Accepted 2013 April 10.  Received 2013 April 10; in original form 2012 November 13}

\pagerange{\pageref{firstpage}--\pageref{lastpage}} \pubyear{2008}

\maketitle

\label{firstpage}

\begin{abstract}
This paper aims at explaining the two phases in the observed specific star formation rate (sSFR), namely 
the high ($>3/Gyr$) values at $z>$2 and the smooth decrease since $z=2$.
In order to do this, we compare to observations the specific 
star formation rate evolution predicted by well calibrated models of chemical evolution
for elliptical and spiral galaxies, using the additional constraints on the mean stellar ages
of these galaxies (at a given mass).
We can conclude that the two phases of the sSFR evolution across cosmic time are 
due to different populations of galaxies. At $z>2$ the contribution comes from spheroids: the progenitors
of present-day massive ellipticals (which
feature the highest sSFR) as well as halos and bulges in spirals (which contribute with
average and lower-than-average sSFR). { In each single galaxy the sSFR decreases rapidly
and the star formation stops in $<$1 Gyr. However the combination of  different generations of ellipticals in formation might result in an apparent lack of strong evolution of the sSFR (averaged over a population) at high redshift.}
 The $z<2$ decrease
is due to the slow evolution of the gas fraction in discs, modulated
by the gas accretion history and regulated by the Schmidt law. The Milky
Way makes no exception to this behaviour.

\end{abstract}


\section{Introduction}

The evolution of the relation between stellar mass and star formation rate is constrained
out to high redshift (e.g., Bell et al., 2005, Elbaz et al., 2007, Noeske et al., 2007, Daddi
et al., 2007, Pannella et al., 2009, Oliver et al., 2010):
at any $z<2$, galaxies form a tight star-forming sequence, with the star formation rate (SFR)
nearly proportional to the stellar mass. In particular,
observations  show that the average specific star formation rate (sSFR,
i.e. the SFR per unit stellar mass) increases by a factor of 20 from z$\sim$0 to z$=$2. What 
happens at z$>$2 is the subject of a lively debate. Whilst
(earlier) studies report an almost constant sSFR ($\sim 2-3/Gyr$) with redshift
 (Gonzalez et al., 2010, Reddy et al. 2012, and references therein), other (more recent) works
support an increase in the sSFR with respect to the $z\sim 2$ value (see, e.g., Stark et al., 2012 and Section 3
of this article). 

The interpretation of the sSFR-redshift evolution is complicated by the fact that
that different populations of galaxies may contribute 
at a given redshift (e.g. Stark et al., 2009), with Renzini (2009)
arguing that the high redshift objects with the largest sSFR might turn into spheroids, whereas
those with lower than average sSFR will evolve as discs. Indeed, a large fraction of $z\sim2$ galaxies show disc-like 
structures (e.g. Cresci et al. 2009, Law et al., 2009, Gnerucci et al., 2011), similarly to the low-redshift
counterparts in the mass-SFR relation. Moreover, despite these
$z\sim 2$ galaxies form stars at a rate of up to a few 100 $M_{\odot}/yr$ (Shapley et al. 2004, 2005
Daddi et al., 2007), there is increasing evidence that such high SFRs are not caused by merger-driven starbursts as it happens
locally (e.g., Rodighiero et al., 2011, Kaviraj et al., 2012). 
A large reservoir of gas, with a supply rate exceeding the gas consumption rate, must be available
to these galaxies in order to maintain their high SFR (e.g. Erb et al., 2006, Daddi et al. 2010) for a relatively
long period. It has been argued that, in 
order to do so, galaxies must behave as self-regulating systems that evolve in nearly steady-state:
the gas accretion (driven by the cosmological accretion rate of the host halo) balances star formation
and outflows. This is shown in several simple (analytic) models
of galaxy formation (e.g. Bouch\'e et al., 2010, Papovich et al., 2011, Dav\'e et al., 2012, Reddy et al., 2012, 
Lilly, et al., 2013, Pipino, et al., in prep.). 
The above-mentioned observations as well as the analysis
performed with simple models, suggest that the galaxies driving
the $z<2$ evolution of the sSFR are disc-dominated. 
How would the Milky-Way fit in this scenario? What is the contribution of earlier
morphologies to the $z<2$ sSFR evolution?

{ If true, the constancy} of the sSFR at higher redshift, instead,
seems to require a conspiracy of many 
physical processes, including accretion, outflows, mergers and the variation of the star formation efficiency with time and/or halo mass (e.g. Weinnmann et al., 2011, Khochfar \& Silk, 2011, but see Krumholz \& Dekel, 2012),
to be explained in the Cold Dark Matter cosmology.
Taken at a face value, these high values of the sSFR reported at $z>2$,
suggest a vigorous star formation history,
which should not differ much from the one inferred for massive spheroids
on the basis of their chemical abundance pattern (e.g. Pipino \& Matteucci, 2004); however,
it has yet to be proven that this is the case.

In order to answer to these questions, 
we tackle the problem of understanding the evolution of the sSFR from the
perspective of chemical evolution. In particular, we make use of models fully calibrated
on $z\sim0$ observables: abundance ratios, ages, SN rates in the Milky Way as well as in local elliptical galaxies.
These models well explain the observed evolution of the mass-metallicity-SFR relation
up to $z\sim3$ (Calura et al., 2009) if one takes into account that populations differing in
morphological type contribute at different epochs, according to the empirical
fact that the average age of stars in elliptical galaxies is larger than that in spirals. 
These models were never tuned to match the observed
sSFR, therefore we can regard the { predicted} sSFR as a genuine result achieved without any 
alteration of the model parameters. 
We aim at 
comparing our results to available observations, 
to test if the empirical trend can be reproduced by the star-forming progenitors of  
present-day galaxies of different morphological types. 
In Section
2 the models are briefly summarized.
In Section 3 we present our results
and discuss them. Conclusions are drawn in Section 4.

\section{The model}

The models adopted in this paper are taken from Pipino \& Matteucci (2004) for the ellipticals, while for a Milky Way like galaxy we make use of the two-infall model of Chiappini et al. (1997, 2001).
In brief, in all these models the gas accretion is regulated by an assumed 
infall law. 
{ Such a law has an exponential form for both ellipticals and spirals:
\begin{equation}
\dot M_{acc} \propto e^{-t/\tau}
\end{equation}
 where $\tau$ is the time scale for the infall (details can be found in the original papers).}

Ellipticals can also eject gas (and stop star
formation) through a supernova-driven galactic wind, if the self-consistently derived
energy budget exceeds the binding energy of the gas.

{ The star formation law is the Schmidt (1959), where the SFR is proportional to some power of the gas density:
\begin{equation}
SFR=\nu \sigma_{gas}^{k}
\end{equation}
where $\nu$, expressed in unit of $time^{-1}$, is defined as the efficiency of star formation, namely the SFR per unit mass of gas, and it should not be confused with the sSFR.
We adopt $\sigma=\sigma_{gas}$ (surface gas density) and the exponent  $k=1.5$ for spirals; whereas
for ellipticals $\sigma$ is the volume gas density, which implies $k=1$ (Kennicutt, 1998).}

The input values for relevant parameters (star formation efficiency, infall timescale, radius) are 
set to those in Calura et al. (2009, Tables 1 and 2), where the reader can find also the 
descriptions of the star formation histories of these systems. In brief
the Milky Way-like (massive) spiral has a central
surface density $\Sigma_0 = 500\, (2000) \rm M_{\odot} pc^{-2}$, a scale length of 3.5 (5) kpc
and a star formation efficiency $\nu = 1$ (2)/Gyr. The infall timescale is $\tau \sim 8$Gyr at the solar
radius in the Milky Way spiral and increasing outward. 
The low (high) mass elliptical model has an effective radius $R_{eff}=$1 (3) kpc and a star
formation efficiency $\nu=$3 (12)/Gyr. The infall timescales are shorter (0.5-0.4 Gyr) than
in the spiral model. 
We adopt a Scalo (1986) IMF for the spirals, and a Salpeter (1955) IMF for ellipticals.
{ However}, as long as we are interested in the sSFR, the IMF effects basically cancel out.
The present-day stellar masses are $\sim 10^{10}M_\odot$ and $\sim 10^{11}M_\odot$
for the low and high mass elliptical, respectively. The Milky-Way has a present-day
stellar mass of $\sim 10^{10.2}M_\odot$, whereas the massive spiral has a mass
of $\sim 10^{11}M_\odot$.

Chemical evolution simulations of elliptical galaxies have been calibrated to reproduce
the observed mass-metallicity and mass-[$\alpha$/Fe] relation in the stars
of elliptical galaxies (e.g., Worthey et al 1992; Trager et al 2000; Thomas et al 2010, Nelan et al. 2005, Bernardi et al., 2006, Graves et al. 2007), under the assumption that both
the star formation timescale and the infall timescale decrease as a function of galaxy mass { (downsizing in star formation and mass assembly)}.
These models also reproduce the observed chemical pattern in the gas of elliptical galaxies
at both low (Pipino \& Matteucci, 2011) and high (e.g. Matteucci \& Pipino, 2001, 
Pipino et al., 2011) redshift. In particular, Pipino et al. (2011) showed that while
relatively small mass ellipticals well reproduce the chemical properties
of well studied $z\sim 3$ Lyman Break galaxies, the most massive galaxies feature
an abundance pattern and the dust content similar to those observed in the highest
($z>6$) QSO hosts.
In this paper, we model elliptical galaxies as single-zone entities of radius 10 $R_{eff}$.

{ As far as spiral galaxies are concerned,  we adopt the above described Milky Way model and a model suitable for a more massive spiral, such as M 101, with the assumption that larger discs evolve faster than smaller ones (e.g. Boissier et al., 2001), 
in analogy with ellipticals and in agreement with observations, while
still being assembled in an inside-out fashion (Yin et al., 2009, Marcon-Uchida \& al. 2010). }
In this paper, the model predictions will always refer to the quantities integrated over the inner 10 kpc.

Our models reproduce the well-established downsizing character of disc galaxies and spheroids, i.e. 
the shorter duration of the star formation in larger galaxies (Calura et al., 09; Pipino et al., 2011).
Whilst single star formation histories predicted by the models have been shown in the original papers, this is the first time that we show the predicted sSFR for our models.

\section{Results \& Discussion}

\subsection{Spirals and the SFR-mass relation at $z<2$}

\begin{figure}
\begin{center}
\includegraphics[width=3in]{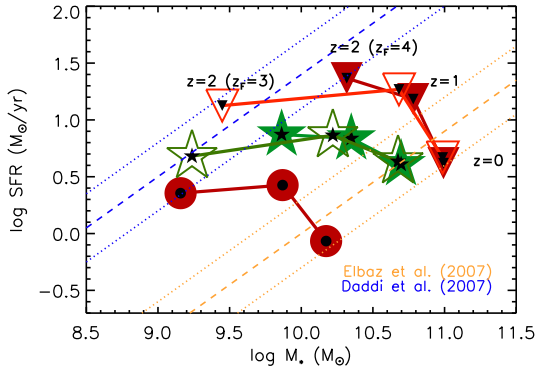}
\caption{The evolution of the Milky Way (filled red circles) and the massive spiral model (stars and triangles) in the SFR-mass plane, as if they were ``observed'' at z=2,1 and 0, respectively.
For the high mass spiral model we show a case in which the infall timescale
is as in Calura et al. (2009, triangles) and one in which it is 3 times longer (stars). The scatter in the observed relation at a given
epoch can be explained by a small scatter in the formation redshift ($z_F$) of these galaxies. In particular, for each of the two model for the massive spiral, we show the cases
with $z_F=3$ (empty symbols) and 4 (filled symbols).
The relations observed at redshift 2 
(Daddi et al., 2007, blue dashed line) and 0 (from Elbaz et al., 2007, yellow dashed line) are shown
for reference. Dotted lines give the $\sim$0.3 dex scatter
of the observed relations around the mean. \label{fig2}}
\end{center}
\end{figure}

The  predicted evolution of the Milky Way model in the 
SFR-mass plane is shown in Fig.~\ref{fig2} by
means of filled red circles.
We also show the massive spiral model. In particular, we show a case in which the infall timescale
is as in Calura et al. (2009, triangles) and one in which it is 3 times longer (stars). 
We compare the model predictions to the
the relations observed at redshift 2 
(Daddi et al., 2007, blue dashed line) and 0 (from Elbaz et al., 2007, yellow dashed line). Dotted lines give the 0.3 dex dispersion
of the observed relations around the mean. 
The star formation history
of the Milky Way model matches the observed evolution of star forming galaxies between z=2 and z=0,
{ being almost a factor of 2 below the average at a given mass, but still within the empirical
dispersion.}
The path of a more massive spiral galaxy is located in the upper part of the SFR-mass plane,
being however very similar to that of the Milky Way.
{ For the generic massive spiral model we did not have any specific constraint on the age. This allow us to show that }
the scatter in the observed relation can be explained by a small scatter in the formation redshift ($z_F$) of these galaxies. In particular, for each of the two models for the massive spiral, we show the cases
with $z_F=3$ (empty symbols) and 4 (filled symbols).
Our models do not incorporate mergers and smooth gas accretion histories are assumed,
therefore we cannot rule out that part of the observed scatter originates from episodic small bursts.

\begin{figure}
\begin{center}
\includegraphics[width=3in]{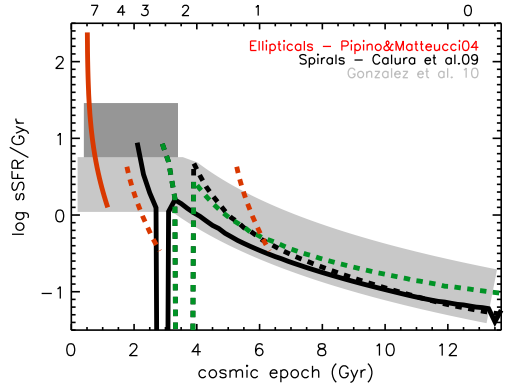}
\caption{Time evolution of the Log sSFR. The light gray area brackets the observed
evolution ($\pm 1 \sigma$) of the sSFR with cosmic time as compiled by Gonzalez et al. (2010,
see also Weinmann et al. 2011) for galaxies of mass $0.2-10.0 \times 10^{10} M_{\odot}$. 
The larger dark gray box qualitatively shows the direction of the change in the
estimated sSFR when dust extinction, correction for emission contamination and a large range of star formation histories
are allowed (see references in the text and Fig.~\ref{fig1bis}).
Spirals (solid: Milky Way; dashed: massive
spiral - a model with a different infall timescale is drawn in green, see text for
details), and ellipticals ({ red \&} dashed: low mass model - { red \&} solid: massive model) are shown.
\label{fig1}}
\end{center}
\end{figure}

In Fig.~\ref{fig1} we show the temporal evolution of the sSFR. The light gray area brackets the observed
evolution ($\pm 1 \sigma$) of the sSFR with cosmic time as compiled by Gonzalez et al. (2010,
see also Weinmann et al. 2011) for galaxies of mass $0.2-10.0 \times 10^{10} M_{\odot}$. 
Let us start by discussing the $z<2-3$ evolution.
At high redshift  the situation is not clear-cut and we defer the discussion
to the next session. 

Since the SFR-mass relation at $z<2$ has a slope close to unity (e.g. Noeske et al., 2007, Elbaz et 
al., 2007, Daddi et al., 2007),
the trend depicted in the figure is a reasonable proxy for the sSFR evolution of star
forming galaxies at any given mass. 

We show our model  spiral galaxies as black lines (solid: Milky Way; dashed: massive
spiral). The formation redshift for the Milky Way model is chosen in order
for the halo (and its associated globular cluster population) to have ages
around $\sim$ 12 Gyrs (De Angeli et al., 2005, Marin-Franch et al., 2009, Thomas
et al., 2011). It is however clear that this model can be used
as a proxy for a ``normal'' spiral, by relaxing the specific constraints on the Milky Way
and allowing for a scatter in the values of $z_F$ as well as the infall timescales.
For clarity purposes, we show the effect of these changes in
the massive spiral model, where the time at which the star formation begins is 
arbitrarily chosen to be $z_F\sim$3.
Two different infall timescales are adopted in the case of the massive spiral: 
as in Calura et al. (2009, black), and 3 times longer (green).

{ In these models the star formation stops at the end of the halo phase due to the
gas density dropping below the star formation threshold and then sets in again after
the second infall has provided enough gas to begin the thin disc formation (Chiappini
et al., 1997). As a consequence there is a short ($<0.5 Gyr)$ period in which the star formation,
and hence the sSFR, drops to zero. This can be seen occurring at about 2.5 Gyr in Fig.~\ref{fig1}.}

The evolution of the mass-SFR relation predicted by our model spirals matches the observations
{ within the scatter} (c.f. Fig.~\ref{fig2}).
Therefore it is not surprising to see that the temporal evolution of the sSFR at $z<2$
is well reproduced. { More specifically, the Milky Way model evolves always below the 
average sSFR at any given time, but still within the observational 0.3 dex scatter.
We do not believe that the offset is significant since there is no guarantee that the Milky Way is
the average galaxy. Furthermore, we avoided any sort of fine tuning. As the dashed black line shows,
a later formation redshift would automatically lead to higher sSFR that predicted
by our best model for the Milky Way at $z=1-2$.
Changing the infall history (the green dashed line) makes the predicted values
higher at $z<1$.
Therefore allowing for the scatter in these properties of otherwise similar galaxies can easily 
populate the gray area}.

\begin{figure}
\begin{center}
\includegraphics[width=3in]{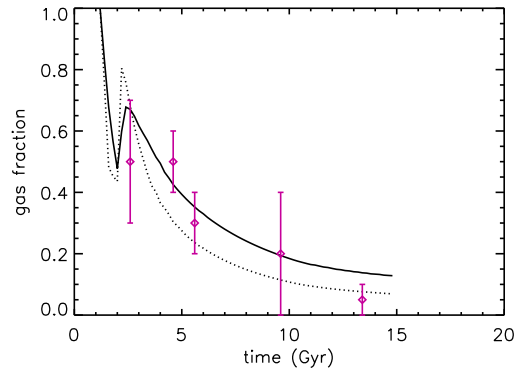}
\caption{Gas fraction evolution predicted for the spirals (solid: Milky Way; dashed: massive
spiral).
Data on the observed molecular gas fraction 
from Tacconi et al. (2010, $z\sim2.5$), Daddi et al. (2010, $z\sim 1.5$), Geach et al. (2011, $z\sim0.4$)
and Leroy et al. (2008, $z=0$) are shown by symbols with error-bars.
\label{fig3}}
\end{center}
\end{figure}

The physical reason behind the agreement with data is that in our models,  the decrease in the star formation
rate over cosmic time is modulated by the infall law.
The empirical infall law that we adopt for the Milky Way is, in turn, very similar to that inferred
from Dark Matter only simulations, constrained to match in mass and reproduce
the relatively quiet history of the Milky Way host halo (Colavitti et al., 2008).
Such an infall law and the assumption of a Schmidt law for the star formation,
induce the evolution in the gas fraction $\mu= \sigma_{gas}/\sigma_{tot}$.
{ The predicted variation in the gas fraction with time is shown in Fig.~\ref{fig3},
where we compare our spiral models to a compilation of observational data
(Tacconi et al., 2010, Daddi et al., 2010, Geach et al., 2010, Leroy et al., 2008). The observed 
galaxies have stellar masses above
$10^{10}M_{\odot}$ and star formation rates such that they are typical star forming objects at their respective
redshift (i.e. they follow the observed relation shown in Fig.~\ref{fig2}, within the scatter). This ensures 
consistency with the other datasets used in this paper and with the stellar masses of the models.}

Since it can be easily shown that $sSFR=\nu \mu^{1.5}\, / (1-\mu)$ (e.g. Reddy et al., 2012)
any change in the gas fraction in the disc will drive a variation in the sSFR.
This empirical finding will be demonstrated 
in more general and theoretical terms in a separate series of papers (see Lilly et al., 2013, and 
Pipino, et al., in prep.). In particular, we anticipate that a smoothly decreasing cosmological accretion rate, taken
from simulations, and the regulating action of the Schmidt (1959) law are sufficient to 
drive the observed variation in the sSFR with cosmic time.

Therefore we conclude that the evolution of the Milky Way in the mass-SFR (equivalently in the sSFR-time)
plane inferred from models tuned to reproduce only its chemical evolutionary history,
{ is consistent with} the average trend of star forming galaxies.
The predicted evolution of the sSFR should not be confused with the notion that 
the SFR in the Milky Way did not change by more of a factor of a few during its
evolution, which is based on the comparison between the current SFR and the past average
star formation rate ($<$SFR$>$)\footnote{Trivially sSFR $\sim$ SFR/($<$SFR$>\cdot$ T), where T is the
cosmic epoch. Therefore the sSFR decreases even if SFR$\simeq<$SFR$>$.}.

We also note that this in agreement with earlier works (Calura \& Matteucci 2003, Vincoletto et al., 2012), 
where we showed that the observed evolution of the luminosity density at $z<2$ is primarily driven disc galaxies.

Going back to Fig.~\ref{fig1} and the $z<2$ evolution of the sSFR,
we note that models for ellipticals are shown with red lines. In particular,
the low mass model is plotted as a dashed line. We show two of such
galaxies, one with redshift of formation of $z\sim1$ (boundary of lowest
ages allowed by spectral analysis of local galaxies, e.g. Thomas et al., 2010) and one at a higher
($z\sim3$) redshift, where it may be observed as a Lyman Break galaxy (Pipino et al., 2011).
The predicted sSFR
agrees with the observations at the same redshift; therefore we cannot exclude that
observational samples at $z\sim 1-2$ contain progenitors of low mass ellipticals. These
galaxies cannot, however, explain the cosmic run of the sSFR down to $z\sim0$.

The analysis of the integrated spectral light tell us that massive ellipticals should
be already in place and/or about to be quenched by $z\sim 2$. Therefore their contribution
to the observational samples made of $z<2$ star forming galaxies must be negligible.

\subsection{The specific star formation rate evolution at $z>$2}

\subsubsection{Model predictions: the sSFR evolution in single elliptical galaxies}

On a qualitative basis, it is clear from Fig~\ref{fig1}, that the high values of the sSFR, averaged over an observational sample, at $z>2$ of $\sim$2-3/Gyr can be easily obtained
with our model ellipticals.
{ In particular, the highest sSFR derived for single galaxies (darker gray box, see discussion 
below) can be traced back to the earliest phases of the formation
of the most massive spheroids, whereas (lower than average) values are typical
of lower mass ellipticals, and possibly the halo/thick disk/bulge components
of spirals.

More specifically, our models predict rather flat star formation histories (c.f. Fig. 1 in
Calura et al., 2009), irrespective of mass, apart from a fast increase in the first 100 Myr.
After the peak, they have a mild decline (Fig.1 in Calura et al., 2009),
certainly shallower than an exponential, until the galactic wind stops the star formation.
We therefore expect our SFR to be qualitatively consistent with Reddy et al. (2012), who find
that strongly declying star formation histories give inconsistent determination of the SFR.
More quantitatively, the low mass elliptical model features SFR$\sim 10 M_{\odot}/yr$, whereas
for the massive one the model predicts SFR$\sim 100 M_{\odot}/yr$. 
While we cannot self-consistently\footnote{We would 
need to convolve our results with either a theoretically motivated
or an observational mass/luminosity function} predict the SFR distribution of galaxies with our chemical evolution simulations,
we can certainly say that these SFRs are reasonable in the light of those observed at $z\sim4-5$, and their distribution 
(Smit et al., 2012).

At the same time, the galactic stellar mass rapidly increases in the models. Therefore, 
for each single galaxy we always predict a decreasing sSFR with time; for instance 
the most massive galaxy model starts from sSFR exceeding 10/Gyr and quickly evolves with time,
much faster than in the case of spirals. 

This predicted behaviour is at variance 
with the face-value interpretation of the those observations (e.g. the compilation
by Gonzalez et al., 2010), that seems to support a plateau in the sSFR. In the following sections
we will show, however, that the plateau might arise from a combination of observational systematics 
and the combination of many populations of otherwise identical ellipticals that form over a suitable redshift range.}

\subsubsection{On the observational results}

{ From the observational
viewpoint the estimated sSFRs at high redshift, are subject to a number of uncertainties, and the average
value, and the very presence of a sSFR plateau at $z>2$ are still being debated. It is worth presenting
a breif discussion, before making a
detailed comparison to our models.}

{ In the first place, it is worth reminding that the observational points
that make the grey area in, e.g., Fig.~1, represents the sSFR averaged over
the galaxies in each observational sample. By inspecting the data, 
we derive that in each observational sample, there is part of the population
with significantly higher than average sSFR, and part with sSFR smaller than 1/Gyr: if the 
proportion between these two populations does not change with redshift this would artificially
create a flat trend in the average sSFR, as we will discuss below.

Moreover, in {those studies (e.g. Gonzalez et al. 2010) that}
use a restricted set of star formation histories  
in their SED-fitting procedure to derive masses \emph{and} SFR\footnote{Which then will be correlated
if, e.g., a constant star formation history template SED is assumed (McLure et al., 2011).} 
and no dust extinction, the small scatter around such mean value is possibly an artifact of this procedure (McLure et al., 2011).}
{ The debate is however still open, as it is not clear whether using a large 
range in star formation histories is physically sound, as, e.g.
the assumption of exponentially declining histories seems to yield inconsistent results 
between the SFR estimated via SED fitting and those derived from UV and IR (Reddy et al., 2012).}

Additionally, an increase in the average sSFR
at $z\sim 4-7$ with respect to the $z\sim2$ value (the ``average sSFR'', lightest gray area, in 
Fig.~\ref{fig1}) has been reported by several very recent works: Bouwens et al. (2012),
move the estimate of the sSFR upward by a factor of 2, due to the revised dust correction. 
The correction for the contamination of nebular emission
lines, that artificially increase the stellar masses derived from SED fitting, yield a similar
up-ward increase in the sSFR estimate (Schaerer \& de Barros, 2010, Stark et al., 2012).
Finally, a further increase in the sSFR estimate is expected if exponentially-increasing
star formation histories\footnote{That is star formation histories that are consistent
with the reported constant sSFR} are adopted (Maraston et al., 2010 { but see Reddy et al., 2012,
Gonzalez et al., 2012).}

In conclusion, the observed average sSFR at $z>4$ may reach values of 5-10/Gyr, considerably larger
than the values observed at $z\sim2$, and in better agreement with our predictions.

\subsubsection{Comparing models to observation}

\begin{figure}
\begin{center}
\includegraphics[width=3in]{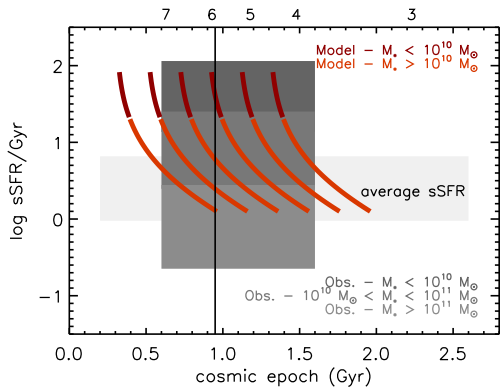}
\caption{Time evolution of the Log sSFR at $z>$2. The light gray area (labeled ``average sSFR'') brackets the observed
evolution ($\pm 1 \sigma$) of the sSFR with cosmic time as compiled by Gonzalez et al. (2010,
see also Weinmann et al. 2011) for galaxies of mass $0.2-10.0 \times 10^{10} M_{\odot}$. 
The three darker gray boxes show the sSFR estimated for
$z\sim 5$ galaxies when dust extinction and a large range of star formation histories
are allowed { (Yabe et al., 2009). In particular, the darkest box (highest sSFR) corresponds to masses 
$<10^{10} M_{\odot}$, whereas the lightest of the three is for galaxies with masses above $10^{11} M_{\odot}$.}
The evolutionary track of the high mass model elliptical is rendered with a darker red in the phases of its
evolution characterized by the stellar mass in the range $0.2-10.0 \times 10^{10} M_{\odot}$.
Such a model is replicated by simply assuming different formation redshifts $z_F$, in such
a way that the onset of the star formation is spaced by 0.2 Gyr. 
The vertical solid line shows how an hypothetical observation at $z\sim 6$ will
read different values of the sSFR of galaxies differing only in $z_F$.
\label{fig1bis}}
\end{center}
\end{figure}

In order to show how some of these effects would contribute 
to increase both the average observed sSFR and
its scatter (see also Fig. 26 in De Barros et al., 2012),
we first construct three dark gray boxes in Fig.~\ref{fig1bis},
whose width along the y-axis is given by 
the 1$\sigma$ scatter around the mean sSFR estimated for
$z\sim 5$ galaxies when dust extinction and a large range of star formation histories
are allowed. In particular, we make use of the values published by Yabe et al. (2009), whose
sample feature a rather large average sSFR. Other observational samples, including those
with dust correction (Bouwens et al., 2012) and emission line corrections (e.g. Schaerer \& de Barros, 2010, Stark et al., 2012),
would be bracketed by the Yabe et al. (2009) values and the Gonzalez et al. (2010) compilation (lightes
gray box labelled as ``average sSFR'').
We repeat the exercise 
for three stellar mass bins
in which we split the Yabe et al. data-set (different shades of dark gray, see Figure caption). 
A strong decrease in the sSFR with stellar mass becomes evident in the data.
{ A caveat that applies to the data-model comparison is that the observations are based on UV-selected
galaxies, therefore low mass galaxies will tend to have the highest sSFRs (e.g. Reddy et al., 2012).}

The width of the box along the abscissa is suitably chosen to bracket
the $z\sim4-7$ redshift range where, according to the latest data (Bouwens et al., 2012, Schaerer \& de Barros, 2010, 
Stark et al., 2012), the systematic uncertainty in the sSFR may amount
to a factor of 2 or larger and the spread in the single galaxy measurements may be significantly
larger than 0.3 dex.

Let us now compare these data to the models.
{ As a first step, mostly for visualization purposes, let us assume that
all galaxies at high redshift are progenitors of massive ellipticals. This is clearly
an extreme assumption, and in reality,
as discussed below, each observational sample is made of galaxies of different morphologies
and masses, distributed according to the mass function, and the observational selection functions. Therefore
the average sSFR should be a weighted average of those predicted by different models. While we
expect that (massive) ellipticals are important contributors at these redshift, a proper
comparison needs a full self-consistent simulation of a cosmologically significant volume.}

The star forming phase of the progenitors of massive galaxies must be short ($<0.6$ Gyr)
to ensure that the chemical composition of their stars is strongly $\alpha$-enhanced.
{ As discussed above, for each single galaxy, the model predicts a quickly declining sSFR. The fastest
decrease is for the most massive ellipticals.}	
Therefore, a galaxy observed at say $z=6$, a few Myr after the 
beginning of the formation, will have $log\,\rm sSFR/Gyr> 1$. 
The same galaxy will have $log\,\rm sSFR/Gyr \sim 0.3$ by $z\sim 5$ 
and it will be passively evolving at $z=4$.
{ These galaxies exhibit for most of the time a high sSFR consistent with that observed. In particular,}
the high mass model elliptical is rendered with a darker red in the phases of its
evolution characterized by the stellar mass in the range $0.2-10.0 \times 10^{10} M_{\odot}$ and
it well matches the data the Yabe et al. (2009) sample.

{ How does this rapid decline compare to the seemingly constant sSFR with redhift?
This might be explained by the fact that the formation of ellipticals
of a given mass seems to be rather synchronized (e.g. Bower et al., 1992, Thomas et al., 2010). Therefore,}
in Fig.~\ref{fig1bis}, we show several evolutionary
tracks created by assuming that the formation redshift of (otherwise identical) high mass models 
for elliptical galaxies are arbitrarily spaced by 0.2 Gyr.
A $\delta z_F\sim1$ Gyr scatter in their formation epoch is fully consistent with classic arguments based
on the scatter of the color-magnitude relation of local ellipticals (Bower et al., 1992).
The assumed redshift of formation are also fully consistent with
spectral analysis of local massive ellipticals, which yields ages of at least 8 Gyr (e.g. Thomas
et al., 2010).
{ The flat probability distribution in $z_F$, instead, is just an assumption for illustration
purposes, and probably does not hold in the real Universe, and does not take into
account the any possible diffence in the comoving number density of these objects
at different redshifts.}

This exercise helps visualizing the fact that, otherwise similar galaxies, differing only in $z_F$ and observed
at, e.g., $z=6$ (vertical line in Fig.~\ref{fig1bis}), will display a range of sSFR values.
That is, if  one allows for a minimal scatter in the formation epoch of these objects,
at each redshift younger - corresponding to higher sSFR - and older (lower
sSFR) objects would coexist. If the probability of formation is uniformly spread over
the redshift range $z=4-10$, and the progenitors of the massive spheroids
dominate the general population of star forming galaxies and behave as in our
models,  then it is easy to see that at any given redshift, an observational sample
will have the same mixture of old and young spheroids, yielding always
an ``average'' sSFR of $\sim 2-3$/Gyr, namely the values around which our model galaxies
spend most of their life. A series of independent generations of galaxies
has been also suggested by the comparison between the evolution in the
the UV luminosity function and the mass-UV luminosity relation 
of high redshift objects (Stark et al., 2009).

\begin{figure}
\begin{center}
\includegraphics[width=3in]{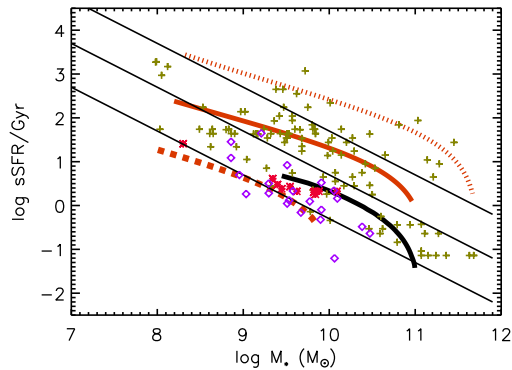}
\caption{sSFR evolution as a function of the stellar mass at the time of observations. Red thick lines
are model ellipticals (solid: massive; dashed: low mass; dotted: additional case with $\sim 10^{12}M_{\odot}$ final 
stellar mass).
For comparison, the massive spiral model is shown by a thick (dark) solid line.
The black thin lines are the relations expected for constant star formation histories
(500, 50, 5 $M_{\odot}/yr$ from top to bottom). Data
from Yabe et al. (2009 - crosses), Gonzalez et al. (2010, asterisks), McLure et al. (2011,
lozenges - converted to Salpeter (1955) IMF).
\label{fig1ter}}
\end{center}
\end{figure}

The situation can be more complex than this, as progenitors of galaxies
with different final stellar masses may contribute to the observations conducted at
a given redshift. 
{ From a qualitative point of view, we can make one futher step and
imagine that the entire population of star forming galaxies is made by two
sub-populations: one represented by our low mass elliptical model, forming
stars at about $\sim 10 M_{\odot}$/yr and with (relatively speaking) lower sSFR, and another
represented by the high mass elliptical model, with higher (specific) star formation rates
(c.f. Fig.~\ref{fig1ter} discussed below).
Since the number densities of objects
with star formation rates $\sim 10 M_{\odot}$/yr is 0.8 dex higher than that of
galaxies forming stars at $\sim 100 M_{\odot}$/yr (Smit et al., 2012), and using
the sSFR predicted by our models, we can roughly estimate that the average sSFR over such a mixed
population will be $\sim 3$/Gyr.}

{ Moreover, the decrease of the number density of highly star forming objects at $z>5$ (Smit et al., 2012),
tells us that we cannot assume a uniform formation probability for the massive
ellipticals at $z>4$. It may decrease with time, leaving us with the conclusion that at the highest
redshifts the smaller ellipticals 
contribute most to the observed sSFR.
Mergers - not included in our models - may then make some of these galaxies create a more massive one.
Whilst it is now well established that late time merger-induced star formation of (Pipino \& Matteucci, 2006) and also
an excess of dry-mergers of ellipticals (Pipino \& Matteucci, 2008, Pipino et al., 2009), do not
reproduce the abundance pattern of present-day ellipticals, less clear is the difference
of mergers and \emph{revised monolithic} models in the high redshift high star forming systems.}

The main massage of the simple exercise discussed above is that we showed that 
achieving a plateau (if any) in the sSFR-time evolution from the point of view of our models 
is relatively easy, once reasonable constraints on $z_F$ and $\delta z_F$ are assumed
{ even if the sSFR steeply declines with time for each single galaxy model.}
Hence it should be straightforward for any other kind
of numerical simulation, once it features the star formation histories for elliptical
galaxies similar to the ones predicted by our model. 
The non trivial task is, instead, to achieve the star formation histories
for ellipticals with suitably high sSFR (and high [$\alpha$/Fe] rations in the stars) from first principles
(Pipino et al., 2009, Sakstein et al., 2011).

Finally, our model for elliptical galaxies suggests that the galaxies with the highest final
mass always feature the highest sSFR at any time along their evolution. This is shown in
Fig.~\ref{fig1ter}, where we plot
the sSFR evolution as a function of the current stellar mass for our
model ellipticals (red lines - solid: massive; dashed: low mass; dotted: additional massive case
built to match the highest sSFR values reported by observations).
The green crosses are the data
from Yabe et al. (2009, $z=5$); whereas red asterisks are from Gonzalez et al. (2010, $z=7$)
and the purple points are from McLure et al. (2011, $z\sim6$).
The black thin lines are the relation expected for constant star formation histories
(500, 50, 5 $M_{\odot}/yr$ from top to bottom), which are adopted by Yabe et al. and Gonzalez
et al., to simultaneously infer masses and star formation rates.
Model ellipticals show a trend which is not very different, despite having star formation
rates first increasing and then decreasing with time.

For comparison, the massive spiral model (only disc phase)
is shown by a thick solid line. As clear already from
Fig.~\ref{fig1}, the early disc phase of a spiral has similar sSFR (at a given mass)
of those predicted by the low mass elliptical model.
At a given mass, there is a large scatter in the data, with our high mass
and low mass elliptical's tracks setting the boundaries to the observed values.

\section{Summary and concluding remarks}

We show the sSFR evolution predicted by chemical evolution models fully calibrated on the elemental abundance
pattern of present day ellipticals and spirals, which also reproduce the chemical properties
of high redshift star forming objects.
{ Our main results can be summarized as follows:}
\begin{itemize}
\item Models for spiral galaxies, where long infall timescales and low star formation efficiencies
      allow a gentle decrease of the gas fraction with time, 
      reproduce the evolution in the SFR-mass relation from z=2 to z=0. In particular, they reproduce the
     observed factor of $\sim 20$
      decrease in the sSFR since $z=2$.

\item The Milky Way makes not exception to this behavior, evolving just above
the average relation (but within the observational scatter).

\item The halo and thick disc phase build-up implies high sSFR values, compatible with those observed at $z>2$.

\item The early phases in models for massive ellipticals, which require strong burst to explain the high
[$\alpha$/Fe] ratios
in their stars, explain galaxies  at $z>2$
with sSFR well above the observational average value.

\item Low mass ellipticals, whose star formation efficiencies (and hence sSFR) are lower than those of massive ellipticals,
      may contribute to the lower than average sSFR observed at $z=1-3$ as well to the ``transition'' phase. Given the similarities in ages and chemical abundance pattern { with ellipticals}, we argue that bulges of spirals, not explicitly studied here, may also contribute.

\item { Whilst in each single galaxy the sSFR decreases rapidly
and the star formation stops in $<$1 Gyr, the combination of  different generations of ellipticals of different mass in formation might result in an apparent lack of strong evolution of the sSFR (average over a population) at high redshift.}

\end{itemize}

We thus conclude that differences in ages, chemical composition and star formation timescales between
spheroid-dominated and disc galaxies, may
coherently explain the two phases of the sSFR evolution across cosmic time as
due to different populations of galaxies. At $z>2$ the contribution comes from spheroids: the progenitors
of present-day massive ellipticals as well as halos and bulges in spirals (which contribute with
average and lower-than-average sSFR).
We argue that a milder evolution of the sSFR
at high redshift may be due to simultaneous observation of different generations of ellipticals in formation.
 The $z<2$ decrease
is due to the slower evolution of the SFR and the gas fraction in discs.
The same physical processes make our suite of models simultaneously reproduce both the mass-metallicity 
and the mass-SFR relations (Mannucci
et al., 2010, Lara-Lopez et al., 2010) observed at different cosmic times, if we consider
different classes of galaxies contributing at different epochs (proto-spheroids at high redshift, discs
at $z<2$, Calura et al., 2009)\footnote{This result has been recently independently confirmed by 
other models (Magrini et al., 2012).}.

\section*{Acknowledgments}
We warmly thank the referee for comments the improved the quality of the paper.


\end{document}